\documentclass[proof]{WileyASNA-v1}

\articletype{Article Type}%

\received{x 2022}
\revised{y 2022}
\accepted{z 2022}

\usepackage{tikz}
\usepackage{graphicx}
\usepackage{bigints}
\usepackage{pgfplots}
\usetikzlibrary{calc}
\pgfplotsset{compat=newest}

\colorlet{myblue}{blue!30}

\begin{document}

\title{The Branch-Cut Cosmology: evidences and open questions}

\author[1,2]{C\'esar A. Zen Vasconcellos*}

\author[3,4]{Peter O. Hess}

\author[5]{Jos\'e de Freitas Pacheco}

\author[1]{Dimiter Hadjimichef}

\author[6]{Benno Bodmann}

\authormark{C.A. Zen Vasconcellos \textsc{et al}}

\address[1]{Instituto de F\'isica, Universidade Federal do Rio Grande do Sul (UFRGS), Porto Alegre, Brazil}

\address[2]{International Center for Relativistic Astrophysics Network (ICRANet), Pescara, Italy}

\address[3]{Universidad Nacional Aut\'onoma de Mexico (UNAM), M\'exico City, M\'exico}

\address[4]{Frankfurt Institute for Advanced Studies (FIAS), Hessen, Germany}

\address[5]{Observatoire de la C\^ote d'Azur, Nice, France}

\address[6]{Universidade Federal de Santa Maria (UFSM), Santa Maria, Brazil}

\corres{*C\'esar A. Zen Vasconcellos. \email{cesarzen@cesarzen.com}}

\abstract[Abstract]{
In this work we present a brief review of the branch-cut cosmology as well as we approach topics not yet investigated. Starting with an already explored topic, the theme of the primordial singularity suppression, we approach essential themes for a better understanding of the scope of branch-cut cosmology. More precisely, based on recent contributions, we advance a better understanding of the branch-cut scenarios as well as essential topics for modern cosmology such as the flatness, homogeneity and horizon problems from the classical point of view of the branch-cut cosmology.}


\maketitle

\section{Branch Cut Cosmology: a brief review}

The main proposal of the branch-cutting model is, --- in addition to suppressing the primordial singularity ---, to consistently propose a contraction phase of the universe, before the big bang explosion and to restore the physical conditions of the expansion period of the universe, without introducing any special feature such as inflation or bouncing, based in turn just on an evolutionary
global topological phase transition description of the universe\footnote{We recently found an article that, although following a very different line of investigation, also seeks to describe the origin of the universe through a topological phase transition. An interesting point of contact of this proposition with the branch-cut cosmology, is that the author describes the birth of the universe using a global topological phase transition with a complex manifold where the time, $\tau$, is considered as a complex variable. However, before the big bang, differently from the branch-cut cosmology it is a purely imaginary variable so that the space can be considered as  Euclidean~\citep{Bellini}.}.

In this sense, the primary idea of branch-cut cosmology is to offer a description of the early evolutionary universe
without resorting to an {\it ad hoc} mechanism that assumes an entire universe created out of essentially nothing, based on the 
``materialization'' of negative energy states, assuming the energy of the gravitational field could become negative without limits~\cite{Guth2004}. 
Or, as proposed also in an {\it ad hoc} way, in the original version of the bouncing model, a non-singular bounce mechanism
that occurs at densities well below the Planck scale where quantum gravity effects are small~\cite{Ijjas2014,Ijjas2018,Ijjas2019}.

The branch-cut classical model of the universe replaces the cosmological singularity with a finely-tuned region, --- purely geometric in nature, which occurs, in the classical version, for densities well below the Planck density scale, where quantum gravity effects may be disregarded  ---,  that allows the smooth transition between the contraction and expansion phases of the universe, thus enabling the resolution of fundamental problems of cosmology~\citep{Zen2022}.  However, as shown recently, the impossibility of packaging energy and entropy according to the Bekenstein criterion~\citep{Bekenstein1981,Bekenstein2003} in a finite size makes the transition phase very peculiar, imposing a topological leap between the two phases or a transition region similar to a wormhole, with space-time shaping itself topologically in the format of a helix-shape like as proposed by the branch-cut cosmology around a branch-point, preserving the fundamental conservation laws during this transition~\citep{Zen2023a}. 

By means of the composition of the multiverse proposal developed by ~\citet{Hawking2018} of a hypothetical set of multiple universes, existing in parallel and the technique of analytical continuation in complex analysis applied to the  FLRW metric~\citep{Friedmann1922, Lemaitre1927, Robertson1935, Walker1937}, 
a new set of Friedmann-type equations for a complexified version of the  $\Lambda$CDM  ($\Lambda\neq 0$) model were obtained~\citep{Zen2020,Zen2021a,Zen2021b}:
\begin{equation}
\Biggl(\frac{\frac{d}{dt} \ln^{-1}[\beta(t)]}{\ln^{-1}[\beta(t)]  } \Biggr)^2   =    \frac{8 \pi G}{3} \rho(t) 
-  \frac{kc^2}{\ln^{-1}[\beta(t)]} + \frac{1}{3} \Lambda \, ;  \label{NFE1} 
\end{equation}
\begin{equation}
\Biggl( \frac{\frac{d^2}{dt^2} \ln^{-1}[\beta(t)]}{\ln^{-1}[\beta(t)] } \Biggr)   =  - \frac{4 \pi G}{3} \Big(\rho(t) + \frac{3}{c^2} p(t) \Bigr)
+  \frac{1}{3} \Lambda 
 ,  \label{NFE2}
\end{equation}
where $\Lambda$ represents the cosmological constant. These equations are
referred  as the first (\ref{NFE1}) and second (\ref{NFE2}) new Friedmann-type  equations analytically continued to the complex plane. The 
analytically continued energy-stress conservation law in the expanding universe is given by:
\begin{equation}
\frac{d }{dt}\rho(t) + 3 \Big(\rho(t) + \frac{p(t)}{c^2}  \Bigr) \Biggl( \frac{\frac{d}{d t} \ln^{-1}[\beta(t)]}{\ln^{-1}[\beta(t)]}\Biggr)  =  0 \, .  \label{ECln}
  \end{equation}%
  This closed set of equations relate a new complex cosmic scale factor, $\ln^{-1}[\beta(t)]$, the energy density, $\rho(t)$, and the pressure, $p(t)$, for a flat, open and closed universe ($k=0, 1, -1$)
(for comparison with the conventional treatment see~\citet{Bazin1965}). The unique feature of this representation of the universe lies in
the restricted superposition of multiverses leading to a single multi-leaf universe in the imaginary domain with a branch-cut type connection around a branch-point while in the real domain the multiverses are disconnected. In this sense, the only domain that allows a plausible physical connected interpretation and allows causality to accomplish is the imaginary domain. 

The complex scale factors $\ln^{-1}[\beta(t)]$, where $\beta(t)$ represents a regularization function\footnote{This notation indicates the reciprocal of $\ln[\beta(t)]$, not its inverse.} allows to define the  analytically continued Hubble rate $H[\beta(t)]$ as
\begin{equation}
H[\beta(t)]  \equiv \Biggl[  \frac{\frac{d}{dt} \ln^{-1}[\beta(t)]}{\ln^{-1}[\beta(t)]} \Biggr]  \, , 
\end{equation}
with
\begin{eqnarray}
  \dot{H}[\beta(t)] & = &  - H^2[\beta(t)] \Biggl( 1  -  \frac{1}{H^2[\beta(t)]} \Biggl[ \frac{\frac{d^2}{dt^2} \ln^{-1}[\beta(t)]}{\ln^{-1}[\beta(t)]} \Biggr]  \Biggr)  \nonumber \\
& \equiv & H^2[\beta(t)] (1 + q_{\rm{ac}}(t)) \,  .
\end{eqnarray}
From this expression, 
\begin{equation}
 q_{\rm{ac}}(t) \equiv  \frac{1}{H^2[\beta(t)]} \Biggl[ \frac{\frac{d^2}{dt^2} \ln^{-1}[\beta(t)]}{\ln^{-1}[\beta(t)]} \Biggr] \, , 
\end{equation}
$q_{\rm{ac}}$ defines the analytically continued deceleration parameter which provides a relationship between the density of the
branch-cut universe and the critical density ($\rho_{\rm{cr}}$), i.e., the density corresponding to  $k = 0$  
for the radiation- and matter-dominated eras~\citep{Zen2020, Zen2021a, Zen2021b}).

Following a similar technical procedure (see~\citet{Zen2020}), we 
arrive at the following complex conjugated Friedmann's-type  equations:
\begin{equation}
 \Biggl(\frac{\frac{d}{dt} \ln^{-1}(\beta^*(t^*))}{\ln^{-1}(\beta^*(t^*))} \Biggr)^2   =     \frac{8 \pi G}{3} \rho^*(t^*)
-  \frac{kc^2}{\ln^{-2}(\beta^*(t^*))} + \frac{1}{3} \Lambda^* ,   \label{NCFE1} 
\end{equation}
and
\begin{equation}
  \Biggl( \frac{\frac{d^2}{dt^{*2}} \ln^{-1}(\beta^*(t^*))}{\ln^{-1}(\beta^*(t^*)) } \Biggr)    =   - \frac{4 \pi G}{3} \Biggl(  \rho^*(t^*)  +  \frac{3}{c^2} p^*(t^*)  \Biggr)
+  \frac{1}{3} \Lambda^*  .   \label{NCFE2}
\end{equation}
The corresponding complex conjugated expression for the energy-stress conservation law in the expanding universe is given by
\begin{equation}
\frac{d \rho^*(t^*)}{d t^*} + 3 \Big(\rho^*(t^*) + \frac{p^*(t^*)}{c^2}  \Bigr) \Biggl( \frac{\frac{d}{d t^*} \ln^{-1}[\beta^*(t^*)]}{\ln^{-1}[\beta^*(t^*)]}\Biggr)  =  0 \, .  \label{CECln}
  \end{equation}%
Similar procedures allow to obtain complex conjugate analytically continued expressions for the Hubble rate, deceleration parameter as well as 
complex and complex conjugated expressions for the analytically continued
Ricci scalar and the Ricci curvature. 

The set of complex equations makes possible the description of the past-present-future evolution of the universe, in which the different Riemann sheets represent space-temporal separations between the different instants of time. In this sense, although outlined in the form of an helix, as the universe evolves, the different instants of time correspond to Riemann sheets that are progressively under construction, thus differentiating the branch-cut model from the Block Universe~\cite{Ellis}, and therefore presenting some similarity with the Evolutionary Block Universe~\cite{Ellis, Ellis2014}.

At this point it is important to reaffirm that these equations are {\it not} a simply direct generalization of the conventional Friedmann's equations based on the real FLRW single-pole metric, nor a simple parametrization of $a(t)$. Due to the non-linearity of the Einstein's equations, such a direct generalization would not be formally consistent. The present formulation as stated earlier, is the outcome of complexifying the FLRW metric and is a result of the solution of a Riemann sum of equations associated to infinitely many poles (in tune with Hawking's assumption of infinite number of primordial universes that occurred simultaneously) arranged along a line in the complex plane with infinitesimal residues (for the details see~\cite{Zen2020, Zen2021a, Zen2021b}). 

In the following, we discuss the singularity, flatness, homogeneity and horizon problems from the classical point of view of the branch-cut cosmology and we make a few comments with other predictions.

\section{Fundamental questions of modern cosmology}

In the following, we discuss some fundamental problems of modern cosmology and propositions of branched cosmology in shedding some light on these questions.

\subsection{The singularity problem and the branch-cut cosmology}

The analytically continued
Ricci scalar, $R_{[\rm{ac}]}  = g_{[\rm{ac}]}^{\mu \nu} R_{[\rm{ac}] \mu \nu}$,
where $R_{[\rm{ac}]\mu \nu}$ defines the analytically continued ([\rm{ac}]) Ricci curvature tensor, becomes
\begin{equation}
R_{[\rm{ac}]}   =  6 \Biggl[  \Biggl(\! \frac{\frac{d^2}{dt^2}{\ln^{-1}[\beta(t)]}}{\ln^{-1}[\beta(t)]}  \Biggr)  +  \Biggl( \frac{\frac{d}{dt}{\ln^{-1}[\beta(t)]}}{\ln^{-1}[\beta(t)]}   \Biggr)^2
  +  \frac{k}{\ln^{-2}(\beta(t))} \Biggr].
\end{equation}
From this expression, we conclude that the new scale factor, $ \ln^{-1}[\beta(t)]$,   the solely dynamical degree of freedom in the analytically continued FLRW metric, 
shapes the curvature of a hypothetical universe with characteristic parameters analytically continued to the complex plane. 
In addition to branch cuts, in the branch-cut cosmology there are `singularities', --- the branch points ---, but at the same time there are multiple points that configure continuous paths in the Riemann sheets. This enables continuous solutions off the primordial singularity, which, in general relativity, are inescapable, allowing this way the analytically continued Ricci scalar curvature do not bend to infinity at the Planck scale, thus eliminating, on the complex plane, the presence of essential singularities.  
For this to happen, the presumption at the level of a local continuity prevails, i.e.,  that there is some neighborhood of the branch point, let's call it $z_0$, close enough although not equal to $z_0$, where one can find a small region around (local patches) where $\ln^{-1}[\beta(t)]$ is single valued and continuous. The cuts in the branch cut, --- shaped by the $\beta(t)$ function ---,  defines the range of  $\ln^{-1}[\beta(t)]$,
 crucial to model the generation of the structures observed today via primordial fluctuations. 
This statement can be visualized by means of the 
representation of the parameter $\ln^{-1}[\beta(t)]$ in polar form with $\beta(t) =  \kappa(t) e^{i n \theta}$, and with the parameter $\kappa(t)$ directly related to the complex analytically  continued Ricci scalar curvature and  to the complex analytically continued {\it radius} of the universe~\cite{Zen2020, Zen2021a, Zen2021b}. Recent calculations based on the Bekenstein criterion have shown that the domain of the beta function is much more extensive than the Planck dimensions, allowing the passage through a kind of `classical tunneling' from the primordial universe, nested to ours, of primordial cosmological matter~\citep{Zen2023a}.

\begin{figure*}[hbpt]
\centering
\begin{minipage}{38pc}
  \includegraphics[width=38mm,height=40mm]{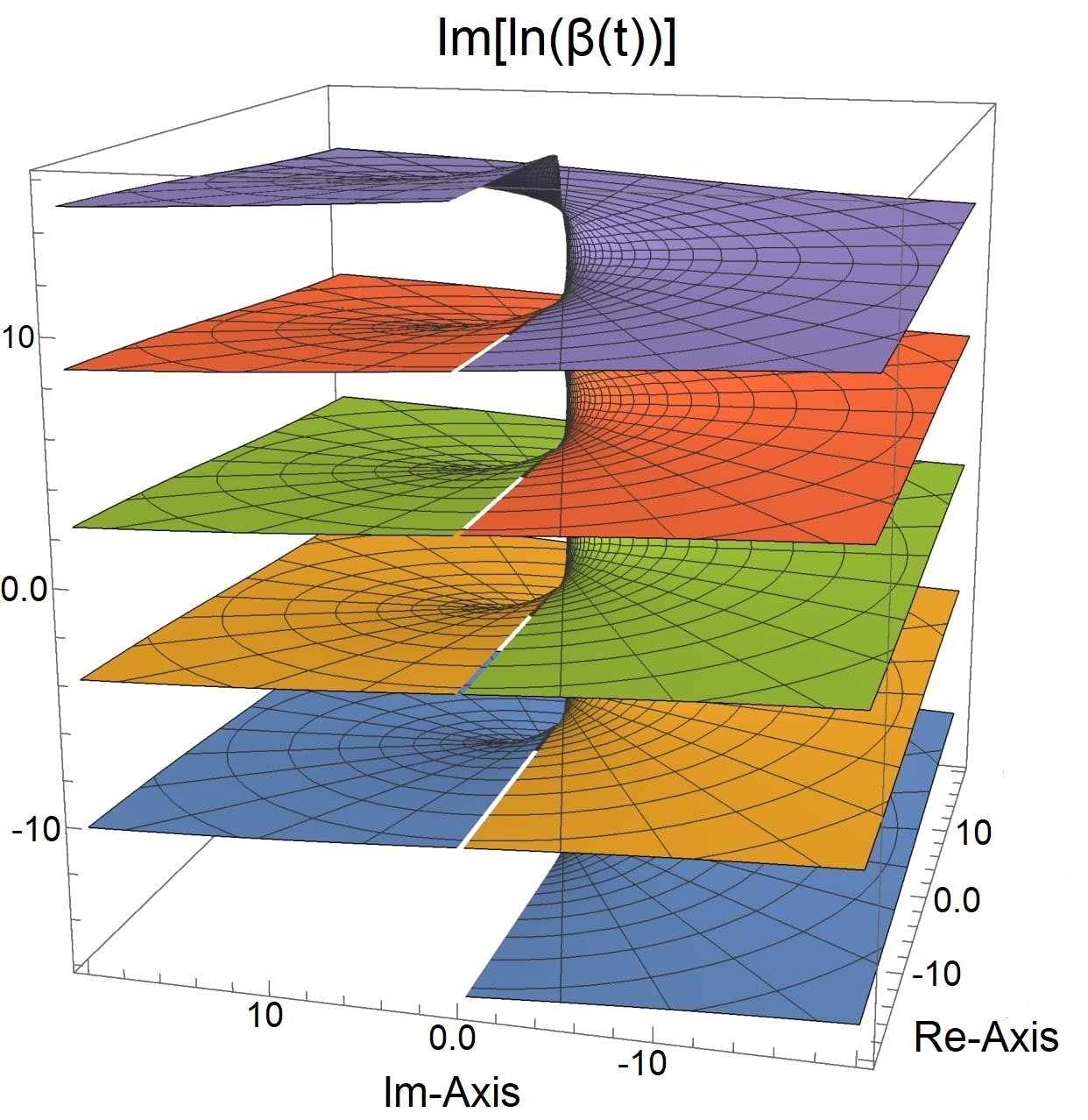} 
\includegraphics[width=38mm,height=40mm]{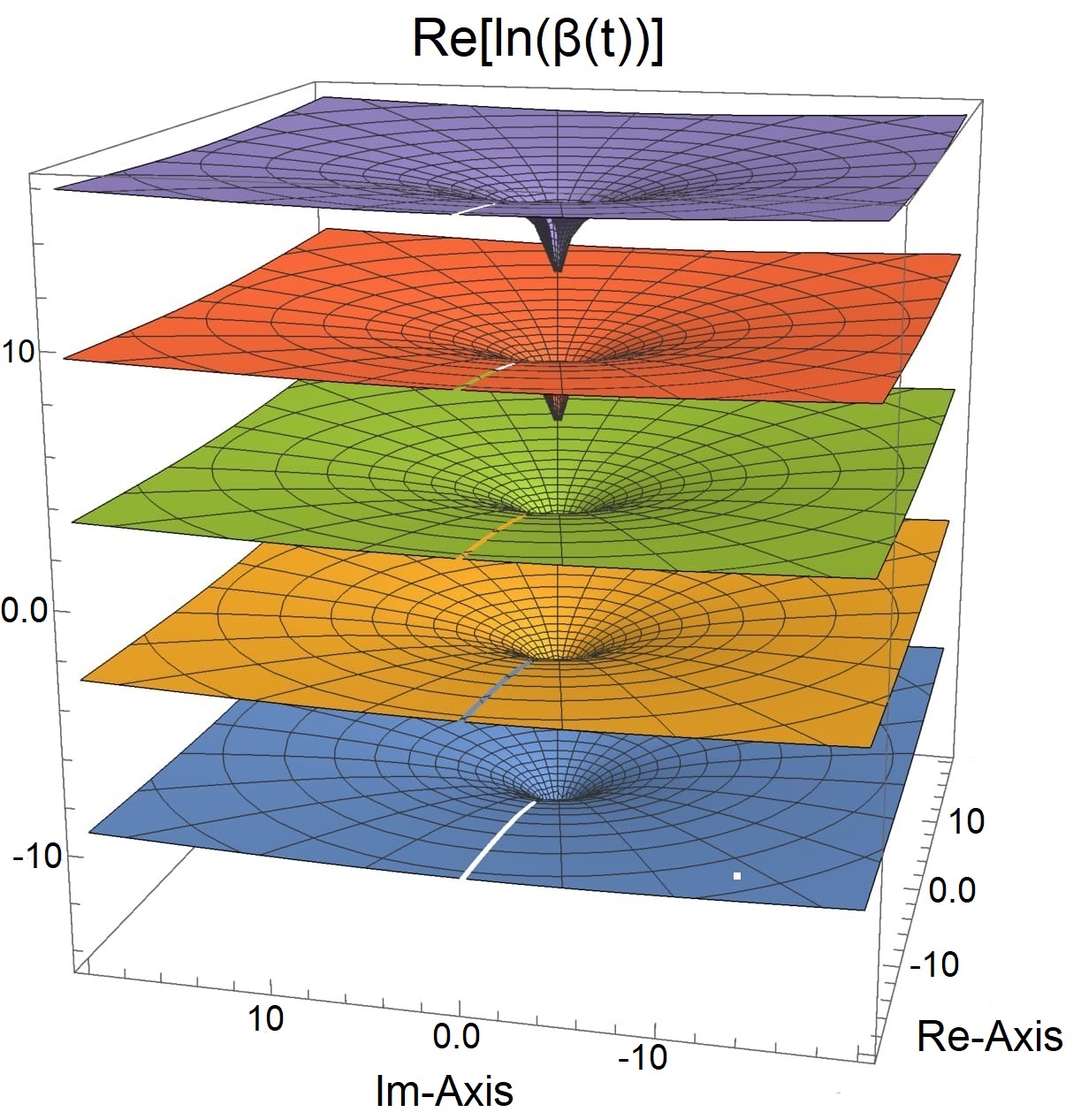}  
   \includegraphics[width=38mm,height=40mm]{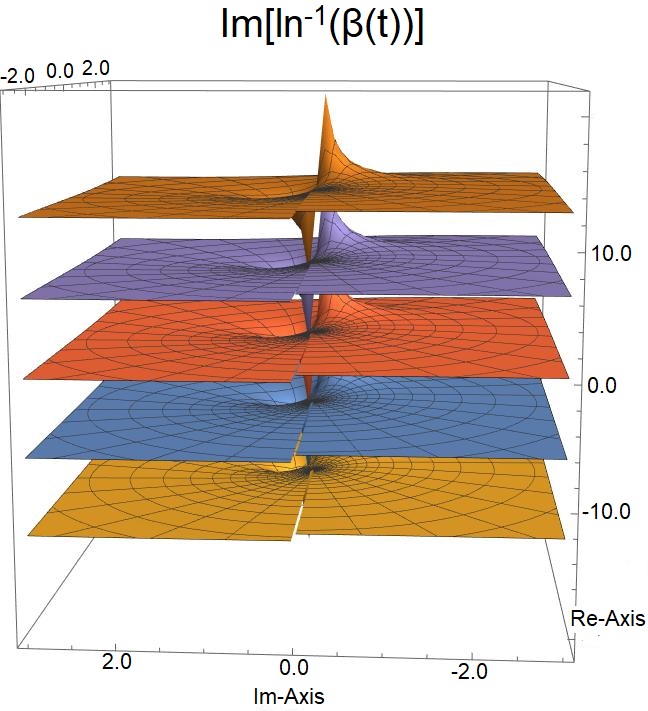}
\includegraphics[width=38mm,height=40mm]{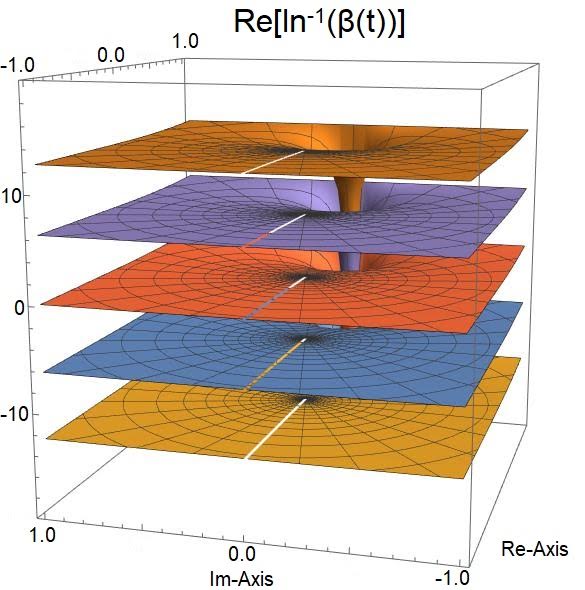}         
\caption{\label{fig2}On the left, the characteristic plots of the Riemann surface $R$ associated with the imaginary and the real parts of the function $\ln[\beta(t)]$ that describes the scaling in time of the branch-cut universe (the reciprocal of $\ln^{-1}[\beta(t)]$). On the extreme left, the plot of the imaginary part shows {\it connected glued domains}: the various branches of the function are {\it glued} along the copies of each upper half plane with their copies on the corresponding lower half plane.  Each two copies can be visualized as two levels of a {\it continuously spiralling parking garage}, from ``level" $\ln z = \ln \kappa(r) + i \theta $ for instance to the ``level" $\ln z = \ln \kappa(r) + i (\theta + 2 \pi) $ or to the ``level $ \ln z = \ln \kappa(r) + i (\theta - 2 \pi) $, and so on. As a final result we have a connected Riemann surface with infinitely many ``levels",  $ \ln z = \ln \kappa + i (\theta \pm 2 n \pi) $, extending clockwise or counterclockwise both upward and downward. For simplicity, the design is limited to a few Riemann sheets. The transition region between the negative and positive sectors corresponds to the predominantly descriptive domains of quantum mechanics and general relativity.  
The figure next to the previous one on the left represents 
the real part of $\ln[\beta(t)]$, with $\tau = |\beta (t)| = \sqrt{\kappa^2_{x} + \tau^2_{y}}$ decomposed in two components in the form $\kappa = (\kappa_{x},\kappa_{y})$ (in a temporal scale of billions of years) and shows a set of multiverses. This procedure allows obtaining complex solutions of Friedmann's-type integral equations  of an evolutionary universe in which the spacetime fabric develops continuously along Riemann sheets that circumvent the branch-cut, thus avoiding discontinuations of the general relativity equations. In short, the corresponding solutions describe a branch cut universe, with a cut from $-\chi(t)$ to $\chi(t)$,  which can be thought as stressed before as a sum of infinity single-poles arranged along a line in the complex plane with infinitesimal residues.  The figures on the right show the characteristic plots of the Riemann surface $R$ associated to the imaginary and real parts of the scale factor of the analytically continued FLRW metric $\ln^{-1}[\beta(t)]$.}
\end{minipage}
\end{figure*}

This representation allows to map the behavior of the 
$\ln^{-1}[\beta(t)]$ parameter in terms of level curves that describe the slope and variations of a hypothetical topological contour.  In a visualization of $\ln[\beta(t)]$ shown in~\cite{Zen2020, Zen2021a, Zen2021b}, 
--- the reciprocal of the new complex scale factor which represents  a linear {\it scaling factor in
time}, bringing to time a complex nature, with an imaginary component\footnote{As historically stated by~\cite{Minkowski1915}, in general relativity and by~\cite{Matsubara1955}, in statistical mechanics.} ---,
the Riemann surface describes a spiral curve around a vertical line corresponding to the origin of the complex plane (Figure  (\ref{fig2})).  
We extended the graphical representation of the previous article showing the real part of this function, where some of the multiverses are represented (Figure  (\ref{fig2})). Evidently, the presence of multiverses in the real part is associated with the peculiar characteristics of the proposal in which the superposed multiple
universes as stressed before correspond to a convenient mathematical device. The representation of the real part of the new scale factor does not contain 
Riemann sheets,
the analytical varieties of complex dimension and therefore the `multi-universes' representation is still present, although with their singularities infinitesimally separated. In complex analysis, contour integrals are made in the complex sector, simply by summing the values of the complex residues inside the contour. 
In Figure (\ref{fig2}) we plot the imaginary and real parts of $\ln^{-1}[\beta(t)]$ for which similar conclusions to the previous case can be drawn.
The actual surfaces of 
the images shown in Figure (\ref{fig2})
extends arbitrarily far both horizontally and vertically the representation sector. 

In short, the present treatment presents a solution for the presence of singularities in general relativity where the Riemann curvature tensor for instance is replaced by an equivalent tensor continued to the imaginary sector that behaves continuously in the real domain and is complemented on the imaginary axis by discontinuity jumps of $2\pi$.

\subsection{Inflation}

Alan Guth has suggested that our universe is product of {\it inflation}~\cite{Guth2004, Guth1981}, an approach based on the existence of states with negative pressure, which effects can be seen in the  Friedmann equations~\cite{Friedmann1922}
combined with the physics of a scalar field, the  {\it inflaton field}. The corresponding
energy-momentum tensor of the {\it inflaton field} $\phi$ is defined as~\citep{Guth2004, Guth1981} 
\begin{equation}
T^{\mu \nu}  = \partial^{\mu} \phi \partial^{\nu} \phi - g^{\mu \nu} \Bigl[ \frac{1}{2} \partial_ {\lambda} \phi \partial^{\lambda} \phi + V(\phi)\Bigr].
\end{equation}
The energy density and pressure in the inflation model are then given by~\citep{Guth2004}
\begin{eqnarray}
\rho(t) & = & T^{00} = \frac{1}{2} \dot{\phi}^2 + \frac{1}{2} \Bigl(\nabla_i \phi  \Bigr)^2 + V(\phi)\,, \nonumber \\
p(t) & = & \frac{1}{3} \sum_{i=1}^3 T_{ii} = \frac{1}{2} \dot{\phi}^2 
 - \frac{1}{6} \Bigl(\nabla_i \phi  \Bigr)^2 - V(\phi).
 \end{eqnarray}
Here $\rho(t)$ is the energy density, $p(t)$ is the pressure, and $V(\phi)$ is the potential energy density. Accordingly, any state dominated by the potential energy of a scalar field will have negative pressure. As a fundamental result, Friedmann's equations then indicate that a negative pressure may contribute to the acceleration of the expansion of the universe, emulating this way a {\it ``repulsive'' form of gravity}. 
The physical state in the original version of the model involved the so-called {\it inflaton} scalar field around a local minimum, a false vacuum, as it behaves, albeit temporarily, as the lowest possible energy density state. For most inflationary models the energy density of the universe $\rho(t)$ is approximately constant leading to exponential expansions of the scale factor $a(t) \sim e^{\chi t}$ (late-time asymptotic behavior), with $\chi = $ const. Thus, the scale factor of evolution of the universe, grows exponentially during the inflation period, causing the quantum fluctuations of a hypothetical scalar {\it inflaton} field to be stretched to macroscopic scales; when leaving the horizon, these fluctuations are frozen. In the later stages of the domain of radiation and matter, these fluctuations re-enter the horizon and thus enables the initial conditions for the formation of structures.  These conditions would lead in the classical case to a completely stable state due to the separation potential barrier from lower energy states. And in the quantum case, --- contrary to expectations of a tunneling process which could cause possibly the ending of inflation ---, it was found that ``the randomness of bubble formation when the false vacuum decayed would produce disastrously (sic) large inhomogeneities"~\citep{Guth2004}. In  Guth's own words~\citep{Guth2004}, a ``graceful exit'' problem was solved by the invention of the new inflationary universe model by~\cite{Linde1982} and~\cite{Albrecht1982}, in which inflation is driven by a scalar field perched on a plateau of the potential energy diagram.

Inflation allows solving at once the {\it monopoles}, {\it horizon} and {\it flatness } problems, among others in the Standard Big Bang (SBB) cosmology.
Guth's original model, despite controversies, still remains as an important component of modern cosmology and has inspired countless authors, as Andrei Linde, Paul Steinhardt, Andy Albrecht and others (see for example~\cite{Martin2014}), as well as a plethora of models in subsequent years to overcome inconsistencies or unexplained features of SSB, despite this theoretical framework being still one of the most accepted model describing the central features of the observed universe. 
Most of these inflation models, --- introduced in an ad hoc way similarly to the original version of the inflationary mechanism, as well as based on general relativity ---, differ fundamentally from each other in terms of the shape of the potential $V(\phi)$ (see, for example, \cite{ Martin2014, Pacheco2022}). 

In this respect, the Cosmic Microwave Background (CMB) measurements by the Planck satellite~\citep{Planck}, has offered an unprecedented opportunity to constrain the inflationary theory, particularly with regard to the scalar spectrum and spectral index of the early universe. These measurements resulted in a certain degree of ordering of the numerous proposals presented so far and at the same time provided researchers with additional constrains for the inflation realization potential $V(\phi)$, regarding in special the spectra
of scalar and tensor fluctuations of the primordial universe, responsible for generating the structures observed today (the scalar spectrum) and the cosmological background of gravitational waves (the tensor spectrum). 

\subsection{The Flatness Problem, inflation and the FLRW cosmology}

 Among unfathomable features of the universe, some stand out. For example, according to general relativity, mass bends space and time. In this context, a universe like ours would be expected to completely curve on itself, with ``positive'' or ``negative'' curvature (using a conventional three-dimensional image, like a ``ball'' in the first case or a ``saddle'' in the second). 

The spatial geometry of the universe has been recently measured by the Planck Collaboration~\cite{Planck} to be nearly flat with high accuracy\footnote{If $\Omega_0$ is less than 1, the universe is open and will continue to expand forever; if $\Omega_0$ is greater than 1, the universe is closed and may eventually stop its expansion and re-collapse; if $\Omega_0$ is exactly equal to 1, then the universe is flat and contains enough matter to stop its expansion, but probably not enough to re-collapse.}, --- the main reason why the flatness problem is understood as a fine tuning problem ---, 
a puzzle that has no explanation in classical FLRW cosmology, giving rise to the so-called flatness problem. 
The flatness problem concerns the value of the ratio $\Omega_{tot} = \rho_{tot}/\rho_c$, where $\rho_{tot}$ is the average total mass density of the universe (including the vacuum energy associated with the cosmological constant $\Lambda$) and $\rho_c$ is the critical density, the density corresponding to $k = 0$. 

In the original version of the FLRW model, during the
radiation-dominated and matter-dominated eras, from the first Friedmann's equation, the scale factor $a(t)$ has the following time dependence
\begin{equation}
\left\{
\begin{array}{l}
\quad a(t) = \Bigl(\frac{H_0}{2}\Bigr)^{1/2} t^{1/2}  \quad \mbox{(radiation-dominated era)}\, ; \\ \quad \mbox{and} \\ \quad a(t) = \Bigl(\frac{3H_0}{2}\Bigr)^{2/3} t^{2/3}   \quad \mbox{(matter-dominated era)}\, , 
\end{array}
\right.
\end{equation}
with the density of the universe $\rho(t)$ and critical density $\rho_c(t)$ (for $k=0$) given by:
\begin{equation}
 \rho(t)  \! = \! \frac{3}{8 \pi G} \Biggl\{ \! \Biggl(\frac{\dot{a}(t)}{a(t)} \Biggr)^2 \! \! + \frac{k}{a^2(t)} \! \Biggr\} \, ,
\! \quad \! \mbox{and} \! \quad \! \rho_c(t)  \! = \! \frac{3}{8 \pi G} \Biggl(\frac{\dot{a}(t)}{a(t)} \Biggr)^2 \!.
\end{equation}
From these equations it results:
\begin{eqnarray}
\frac{\rho(t) - \rho_c}{\rho_c} & = & \frac{\rho(t)}{\rho_c} - 1 = \Omega_{tot} - 1 = \frac{k}{a^2(t)} 
\nonumber \\
& = &  \left\{ \begin{array}{l}
  \frac{2k}{H_0}t^{-1} \quad \mbox{(radiation-dominated era)} \\ \\
   \frac{2k}{3H_0}t^{-4/3}  \quad \mbox{(matter-dominated era)}  \\ 
  \end{array} . \right. \label{Flat}
\end{eqnarray}
Thus, in the limit where $t \to 0$, the FLRW cosmology results are inconsistent with the observational predictions.
However, while the extraordinary flatness of the early universe as measured by the Planck Collaboration~\cite{Planck} has no explanation in classical FLRW cosmology, it is a `natural' prediction for inflationary cosmology. During the inflationary period, instead of $\Omega_{tot}$
 being driven away from one as described by FLRW cosmology, it  is driven toward one, with exponential swiftness:
 \begin{equation}
 \Omega_{tot} - 1 \propto e^{-2 H_{inf} t} \quad  \mbox{leading to} \quad \Omega_0 = 1.02 \pm 0.02.,
 \end{equation}
 where $H_{inf}$ is the Hubble parameter during inflation. 
 
 However, as previously mentioned, relevant questionings to this formulation persist, such as resorting to an {\it ad hoc} mechanism that assumes an entire universe created out of essentially nothing, based on the ``materialization'' of negative energy states, assuming the energy of the gravitational field could become negative without limits~\cite{Guth2004}. The same can be said about the standard cosmology that presents some formal and conceptual inconsistencies, as seen above and as we will see below. In this sense, as we intend to demonstrate in the continuation, the branch-cut cosmology presents itself as a descriptive alternative of the evolutionary universe whose potentialities require a deeper approach, which we intend to do in this contribution.

 \subsection{Patch and horizon sizes scaling and causality in the standard cosmology}

 In the conventional FLRW cosmology, the cosmic scale factor $a(t)$ and the Hubble parameter  both characterize as a function of time a smooth and isotropic universe. The cosmic scale factor $a(t)$ is a dimensionless quantity that describes the size variation of a patch of space due to expansion or contraction of the universe and $H(t) = \frac{\dot{a}(t)}{a(t)}$ in turn measures the expansion rate of the universe. Assuming the observable universe corresponds to a patch of space with radius $R(t^{\prime})$ at time $t^{\prime}$, at time $t$  
the corresponding patch size in the standard cosmology is sized up by $a(t)$
\begin{equation}
R(t)  \Rightarrow  \frac{a(t)}{a(t^{\prime})} R(t^{\prime})\, , 
\end{equation} 
and the reciprocal of $H(t)$ measures the Hubble radius or horizon size:
\begin{equation}
     H^{-1}(t)  =  a(t) \int_0^t \frac{dt^{\prime}}{a(t^{\prime})}\, ,  \label{a&H}
\end{equation}
which represents the maximal region that is ``causally connected'' through interactions with light or any other particles.  The horizon size thus defines the maximum distance to be traveled by light in the age of the universe and therefore the maximum observation distance of any event since the big bang. 

In the radiation dominated era, with $a(t) \sim t^{1/2}$, from the FLRW metric~\citep{Friedmann1922, Lemaitre1927, Robertson1935, Walker1937}, since  light propagates along world
lines for which the spacetime interval is zero,
\begin{equation}
ds^2 = c^2 dt^2 - a^2(t) \Bigl(\frac{dr^2}{1 - k r^2}  + r^2 \bigl(d\theta^2 + \sin^2\theta d \phi^2 \bigr)\Bigr) = 0 \, , \label{rd}
\end{equation}
we get
\begin{equation}
H^{-1}(t)  \! = \!  a(t) \!\! \int_0^t \frac{c dt^{\prime}}{a(t^{\prime})} \! = \! a(t) \!\! \int_0^{r_H} \frac{dr}{\sqrt{1 - kr^2}} 
 \! \sim \!  t^{1/2} \!\! \int_0^t \frac{c dt^{\prime}}{t^{\prime 1/2}} \! = \! 2 ct \, . \label{md} 
\end{equation}
For the matter-dominated era, $H^{-1}(t) = 3 ct$. Thus, in standard cosmology, this simple calculation shows that causality could be established only if a signal could propagate at about 2 or 3 times the speed of light, a proposition that clearly contradicts the known laws of physics.

 A key issue in cosmology concerns the geometric scaling properties of the horizon size. To address this topic, we consider the energy-stress conservation law, in the standard approach:
\begin{eqnarray}
&& \frac{1}{\rho(t)}\frac{d}{dt}\rho(t)    +    3 \Biggl( 1 + \frac{p(t)}{c^2\rho(t)}   \Biggr) \frac{1}{a(t)} \frac{da(t)}{dt} = 0 \, ,  \nonumber \\
&& \Rightarrow \quad \frac{d}{dt}\ln  \bigl(\rho(t)\bigr)  +  
3 \Biggl(1 + \frac{p(t)}{c^2\rho(t)}  \Biggr)  \frac{d}{dt} \ln a(t) =  0.  \nonumber \end{eqnarray}
From this equation it results 
\begin{equation}
\int_{\rho_0}^{\rho(t)} d \ln \bigl( \rho(t) \bigr)  =  - 2 \int \Biggl\{ \frac{3}{2}\Biggl(1 + \frac{p(t)}{c^2 \rho(t)}  \Biggr) \Biggr\} d\ln a(t) \, ,
\end{equation}
where
\begin{equation}
 \epsilon(t) \equiv \frac{3}{2} \Biggl(1 + \frac{p(t)}{c^2 \rho(t)}  \Biggr). \label{r0}
\end{equation}
In this equation $\epsilon(t)$ represents the dimensionless {\it thermodynamics connection} between the energy density $\rho(t)$ and the pressure $p(t)$ of a perfect fluid thus enabling the fully description of the equation of state (EoS) of the system. Positive pressure corresponds to $\epsilon > 3/2$,  negative pressure to  $\epsilon < 3/2$ and for a universe
dominated by a cosmological constant, $\epsilon \to 0$. 
In the limit in which the 
dimensionless thermodynamics connection obeys 
$\epsilon(t) \to  \epsilon = $ constant, the integral (\ref{r0}) reduces to\footnote{The value of $\epsilon$ is nearly constant over long epochs, varying rapidly from one constant value to another
when the dominant form of energy changes. For example, in the standard big bang model,
the universe is radiation-dominated (corresponding to $\epsilon = 2$) for the first 50,000 yrs after the big bang and dust-dominated (corresponding to $\epsilon = 3/2$) for the remaining 13.8 Gyrs.}
\begin{eqnarray}
\ln (\rho(t)/\rho_0) & = & - 2  \lim_{\epsilon(t) \to \epsilon} \int \epsilon(t) d \ln a(t) \nonumber \\
& \simeq & \! - 2 \epsilon \ln a(t)  \Rightarrow \ln  \Bigl(\! a(t) \! \Bigr)^{-2\epsilon} 
 \!\!\!\!\!\!\!  \Rightarrow \! \rho(t) \simeq \frac{\rho_0}{a^{2 \epsilon}(t)}. \label{rhoscaling}
\end{eqnarray}
From this equation, when going back in time considering the primordial universe, if $\epsilon$ is positive, the universe's density increases, while for negative $\epsilon$, its density decreases.
From the first Friedmann's equation in the FLRW cosmology, for $k=0$, combined with the scaling behaviour of $\rho(t)$ (see Equation (\ref{rhoscaling})),  we
have
\begin{eqnarray}
H^2 & = & \Biggl(\frac{\dot{a}(t)}{a(t)} \Biggr) = \frac{8\pi G}{3} \rho(t) \sim \frac{8 \pi G}{3} \frac{\rho_0}{a^{2 \epsilon}(t)}  \nonumber \\
& \Rightarrow & H^{-1} = \sqrt{\frac{3}{8 \pi G \rho_0}} a^{\epsilon}(t). \label{scalingH}
\end{eqnarray}
From Equation (\ref{a&H}), we have seen that the curvature $R(t)$ of the universe scales with the parameter $a(t)$ and from Equation (\ref{scalingH}), that the horizon size,
$H^{-1}(t)$ scales as  $a^{\epsilon}(t)$ so their ratio continuously decreases when extrapolating $a(t)$ backwards in time. 
This scaling difference between the curvature of the universe and horizon size 
represents a fundamental limitation of standard cosmology, generating a non-causal behavior of the patch size and the horizon size.

\subsection{Flatness, Horizon, Inhomogeneity, and Smoothness Problems
in the Branch-cut Cosmology}

In the branch-cut cosmology, 
from the first Friedmann's equation, extended to the complex plane, the following expression, for $\Lambda = 0$ and $k \neq 0$ then holds
\begin{equation}
\Biggl( \frac{\frac{d}{dt} \ln^{-1}[\beta(t)]}{\ln^{-1}[\beta(t)]  } \Biggr)^2       =    \frac{8 \pi G \rho(t)}{3} 
 -  \frac{kc^2}{\ln^{-2}[\beta(t)]}  \, ,
 \end{equation}
 or
 \begin{equation}
 \Biggl( \frac{d}{dt} \ln^{-1}[\beta(t)] \Biggr)^2       =    \frac{8 \pi G}{3}\rho(t)  \ln^{-2}[\beta(t)]
 -  kc^2  \, . \label{FE}
\end{equation}
Solving equation (\ref{FE}) we get the solutions:
\begin{equation}
 \ln^{-1}[\beta(t)]  =     \left\{ \begin{array}{l}
  \sqrt{\ln^{-2}[\beta(t_P)]   
+ \frac{1}{ \ln^{2}(\beta_0)} \sqrt{\frac{2 \pi G \rho_0}{3}} \Bigl( t - t_P  \Bigr) 
 } \\  \mbox{(radiation-dominated era)} \\ \\
   \sqrt[2/3]{ \ln^{-3/2}[\beta(t_P)]   
 +  \frac{1}{ \ln^{\, 3/2}(\beta_0)}  \sqrt{6 \pi G \rho_0} \Bigl( t - t_P  \Bigr)
 }  \\  \mbox{(matter-dominated era)}  \\ 
  \end{array}\, . \label{SFE} \right.
\end{equation}
From expression (\ref{FE}), we obtain the following expressions for the density of the universe, $\rho(t)$, and the critical density, $\rho_c(t)$: 
\begin{equation}
\rho(t) = \frac{3}{8 \pi G} \Biggl\{ \Biggl( \frac{\frac{d}{dt} \ln^{-1}[\beta(t)]}{\ln^{-1}[\beta(t)]  } \Biggr)^2         
 +  \frac{kc^2}{\ln^{-2}[\beta(t)]} \Biggr\} \, ,\label{Rho}
\end{equation}
\begin{equation}
\rho_c(t) = \frac{3}{8 \pi G} \Biggl( \frac{\frac{d}{dt} \ln^{-1}[\beta(t)]}{\ln^{-1}[\beta(t)]  } \Biggr)^2    \, . \label{Rhoc}
\end{equation}
Combining (\ref{SFE}), (\ref{Rho}) and  (\ref{Rhoc}) we finally obtain:
\begin{eqnarray}
\frac{\rho(t) - \rho_c}{\rho_c}  =  \Omega_{tot} - 1 
& = & \frac{k}{\Bigl( \ln^{-1}[\beta(t)]\Bigr)^2}  \nonumber \\ & = &  
 \left\{\!  \begin{array}{l}
  \frac{k}{\ln^{-2}[\beta(t_P)]   
+ \frac{\sqrt{2 \pi G \rho_0}}{\sqrt{3}\ln^{2}(\beta_0)}(t - t_P)} 
  \\ \mbox{(radiation-dominated era)} \\
    \frac{k}{\bigl(\ln^{-3/2}[\beta(t_P)]   
 +  \frac{ \sqrt{6 \pi G \rho_0}}{ \ln^{\, 3/2}(\beta_0)} (t - t_P)\bigr)^{4/3}}
    \\ \mbox{(matter-dominated era)}  \\ 
  \end{array} \right.
\end{eqnarray}
This result looks at a first glance quite similar to the corresponding result of the original FLRW model (see Equation (\ref{Flat})). Once again care must be taken in this interpretation.  
%%%%%%%%%%%%%%%%%%%%%%%%%%%%%%%%%%%%%%
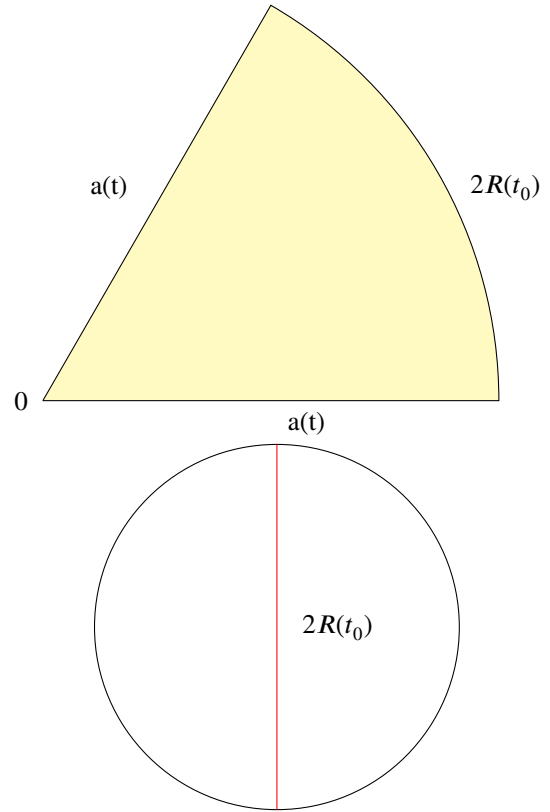
\begin{figure}
\centering
\begin{tikzpicture}
\draw[fill=yellow!30] (0,0) -- (6,0) arc[start angle=0, end angle=60,radius=6cm] -- (0,0);
%\draw (0,0) -- (5.85,1.37);
%\draw (0,0) -- (4.5,3.95);
\filldraw[black] (-0.5,0.0) % circle (2pt)
node[anchor=west]{0};
\filldraw[black] (3.1,-0.3) % circle (2pt)
node[anchor=west]{a(t)};
\filldraw[black] (0.5,2.80) % circle (2pt)
node[anchor=west]{a(t)};
\filldraw[black] (5.5,2.80) % circle (2pt)
node[anchor=west]{$2R(t_0)$};
%\draw[fill=cyan!30] (0,0) -- (0,1.5) arc [start angle=90, delta angle=30, radius=1.5cm] -- (0,0);
% arc (120:360:0.5) ;
\end{tikzpicture}
\hspace{0.8cm}
\begin{tikzpicture}[scale=2] 
\def\L{1.5mm}
\draw[clip] (0,0) circle(1.2);
%\draw (-1,-1) grid[ystep=0,xstep=\L] (1,1);
\draw[red] (0,-1.5)--(0,1.5); % for checking grid go through the center, should be removed
\node[] at (0.4,0) {$2R(t_0)$};
\end{tikzpicture}
\caption{In the upper figure: an arc of a circle containing a portion of space representing schematically the relationship between the patch size and the horizon size,  with the sides of the arc labeled by the scale factor $a(t)$. The curved portion of the arc of circle represents the linear size of a patch of space at the present time $t_0$, represented by $2R(t_0)$. On the below figure: the diameter $2R(t_0)$ of a circle of radius $R(t_0)$, associated to the presently observed universe.} \label{arc}
\end{figure}

%%%%%%%%%%%%%%%%%%%%%%%%%%%%%%%%%%%%%%%%%%%%%%%
In the classical scenario of the branch-cut cosmology, the universe evolves continuously from the negative complex cosmological time sector, prior to a primordial singularity, to the positive one, circumventing continuously a branch-cut, and no primordial singularity occurs in the imaginary sector, only branch-points. The branch-cut universe involves a continuous sum of an infinite number of infinitesimally (originally) separated poles, surrounding a primordial branch-point, arranged along a line in the complex plane with infinitesimal residues.  And similarly to the primordial branch-point singularity, the argument of the resulting analytic function, can be mapped from a single point in the domain to multiple points in the range. 

As stressed before, the flatness problem concerns the value of 
the ratio $\Omega_{tot} = \rho_{tot}(t)/\rho_c(t)$.
The problem of the flatness of the universe is related, additionally, to the  cosmic curvature factor~\citep{Ijjas2014,Ijjas2018,Ijjas2019}, 
a time-dependent and dimensionless quantity that characterize the {\it apparent 
spatial curvature} 
of the universe,
$\Omega_c$, defined in the FLRW cosmology
as 
\begin{equation}
\Omega_c \equiv - \frac{k H^{-2}(t)}{a^2(t)}  \quad \mbox{which scales as} \quad \sim \frac{a^{2\epsilon}(t)}{a^2(t)}.
\end{equation}
Moreover, in the standard cosmology, the horizon problem 
arises exactly because the patch corresponding to the observable universe was never causally connected in the past~\citep{Ijjas2014,Ijjas2018,Ijjas2019}.  As we mention previously, in the present time ($t = t_0$), the patch size, $R(t_0)$, and the horizon size, $H^{-1}(t_0)$, are equal, i.e., $R( t_0) = H^{-1}(t_0)$. In earlier times, the ratio between the horizon size to the patch size decreases monotonically extrapolating back in time as $a(t) \rightarrow 0$ in the form
$a^{\epsilon}(t)/a(t)$:  
\begin{equation}
\frac{H^{-1}(t)}{a(t)} \sim 
\frac{a^{\epsilon}(t)}{a(t)} = a^{\epsilon - 1}(t) \quad
\mbox{and} \quad  \lim_{a(t) \rightarrow 0}{a(t)^{\epsilon - 1}} \rightarrow 0, \label{aepsilona}
\end{equation}
with the horizon size approaching zero faster than the patch size.
According to the CBM measurements, the density and temperature were almost uniform throughout the primordial patch (last CBM surface scattering). Explaining the uniformity of the CMB at length scales greater than the size of the horizon at the last scattering surface grounded on the CBM measurements
and at all previous times fundamentally constitutes the horizon problem~\citep{Ijjas2014,Ijjas2018,Ijjas2019}. Moreover, the CMB measurements also reveal a spectrum of small amplitude density fluctuations, nearly scale-invariant whose explanation constitutes the inhomogeneity problem~\citep{Ijjas2014,Ijjas2018,Ijjas2019}.

As we stressed before, this non-causal behavior of the patch size and the horizon size, --- since their ratio continuously decreases when extrapolating $a(t)$ backwards in time ---, represents one of the fundamental limitations of standard cosmology. Combined with the primordial singularity, where any trace of causality completely disappears, these two factors represent the main roots for solving the problems of the cosmic singularity, horizon, inhomogeneity and flatness of the universe. The 
smoothness problem of the universe is related in turn to the
{\it cosmic anisotropy factor}, $\Omega_a$, a time-dependent, dimensionless quantity that characterizes the apparent anisotropy, due to small temperature fluctuations in the primordial blackbody radiation:
\begin{equation}
\Omega_a \equiv \frac{\sigma^2 H^{-2}}{a^6 (t)} \quad \mbox{which scales as}
\quad \sim \frac{a^{2\epsilon}(t)}{a^6(t)};
\end{equation}
in this case, the patch size approaches zero faster than the horizon size.

The solution of the branch-cut-type cosmology corresponds as we have seen to the reciprocal of a complex  multi-valued function, the natural complex logarithm function $\ln[\beta(t)]$. 
Taking the ratio
\begin{equation}
\frac{\Bigl(\ln^{-1}[\beta(t)]\Bigr)^{2\epsilon}}{\Bigl(\ln^{-1}[\beta(t)]\Bigr)^2} = \ln^{2(1 - \epsilon)}[\beta(t)] \, , 
\end{equation}
and
\begin{equation}
  \lim_{\ln[\beta(t)] \rightarrow 0}{\ln[\beta(t)]^{2(1 - \epsilon)}(t)} \rightarrow   \Biggl\{ \begin{array}{l}
0 \quad for \quad 1 - \epsilon > 0 ; \\
\infty \quad for \quad 1 - \epsilon < 0 .
 \end{array}
\end{equation}

Going from the present to the past of the branch-cut universe, as we return to the beginning of the expansion phase,
as the patch size decreases with a linear dependence on $\ln[\beta(t)]$, light travels through geodesics on each Riemann sheet,  circumventing continuously the branch-cut,  and although the horizon size scale with $\ln^{\epsilon}[\beta(t)]$, the length of the path to be traveled by light compensates for the scaling difference between the patch and horizon sizes. Making an analogy with a continuously spiralling parking garage, everything happens as if two cars left a certain level of the garage at the same time, so that one car, slower, followed a straight ramp while the other car, faster, followed the longer, circular route of the garage, and both arrived at their destination at the same time. 

Under these conditions, causality between the horizon size and the patch size may be achieved through the accumulation of branches in the transition region between the present state of the universe and the past events. 
%In the standard cosmology, the patch corresponding to the observable universe was never causally connected in the past~\citep{Ijjas2014}. 
%In the contraction phase of the branch-cut cosmology, as the patch size decreases with a linear dependence of the complex cosmic scale factor $\ln[\beta(t)]$,
%light travels through geodesics in each Riemann leaf, continuously contouring the branch-cut, and although the horizon size scales with $\ln^{\epsilon}[\beta(t)]$ and the patch size in turn scales with $\ln[\beta(t)] $, the %length of the path the light travels locally around the branch-cut compensates for the difference in scale between the patch and horizon sizes. In these circumstances, the causality between the
%horizon-size and the patch-size can be achieved through the accumulation of branches in the transition region between the current state of the universe and past events. 
This topic, the number of ramifications accumulated to achieve causality was recently addressed in~\citep{Zen2023b}. Ijjas and Steinhardt~\citep{Ijjas2014,Ijjas2018,Ijjas2019} have shown that
quantum fluctuations responsible for temperature variations observed in the Cosmic Microwave Brackground (CBM)
are produced with roughly 60 e-folds of contraction in the FLRW cosmic scale factor $a(t)$ remaining
before the bounce when the horizon size is exponentially large compared to the
Planck scale. 

In a recent calculation we showed that the number of `turns' or Riemann sheets to achieve causality is $\sim 10^{60}$, a number comparable to 60 e-folds of contraction in the bouncing model~\citep{Zen2023b}. More precisely, the e-folds of bouncing contractions are comparable to the running-number or circulation sheets or number of branches in between the present state of the universe and the region where causality is achieved, the cosmic microwave background radiation's {\it surface of last scatter}.

 However, the impossibility of packaging energy and entropy according to the Bekenstein Criterion in a finite size makes the transition phase very peculiar, imposing a topological leap between the two phases or a transition region similar to a wormhole, with space-time shaping itself topologically in the format of a helix-shape like as proposed by branch-cut cosmology around a branch-point. Topological wormholes are said to essentially constitute `tunnels' in spacetime, in which two topological sheets, at the top and at the bottom, are connected via a wormhole. The two topological sheets can represent two different universes or two sections of the same universe which, without the presence of the wormhole, would be separated by a very non-causal large distance. 
 
 Branch-cut cosmology has similarities with wormholes, in the sense the $\ln[\beta(t)]$ function corresponds to a helix-like superposition of cut-planes, the Riemann sheets, with an upper edge cut in the $n$-th plane joined with a lower edge of cut in the ($n + 1$)-th plane. In this sense, the topology of a wormhole is reproduced successively between pairs of upper-edge and lower-edge cut-planes. $\ln[\beta(t)]$ in turn maps an infinite number of  Riemann sheets onto horizontal strips, which represent in the branch-cut cosmology the time evolution of the time-dependent horizon sizes. The patch sizes in turn maps progressively the various branches of the $\ln[\beta(t)]$ function which are {\it glued} along the copies of each upper-half plane with their copies on the corresponding lower-half planes. In the branch-cut cosmology, the cosmic singularity is replaced by a family of Riemann sheets in which the scale factor shrinks to a finite critical size, --- the range of $\ln^{-1}[\beta(t)]$, associated to the cuts in the branch cut, shaped by the $\beta(t)$ function ---, well above the Planck density.

\subsection{Scenarios for the Branch-cut cosmology}

\begin{figure*}[htb]
\centering
\includegraphics[width=70mm,height=45mm]{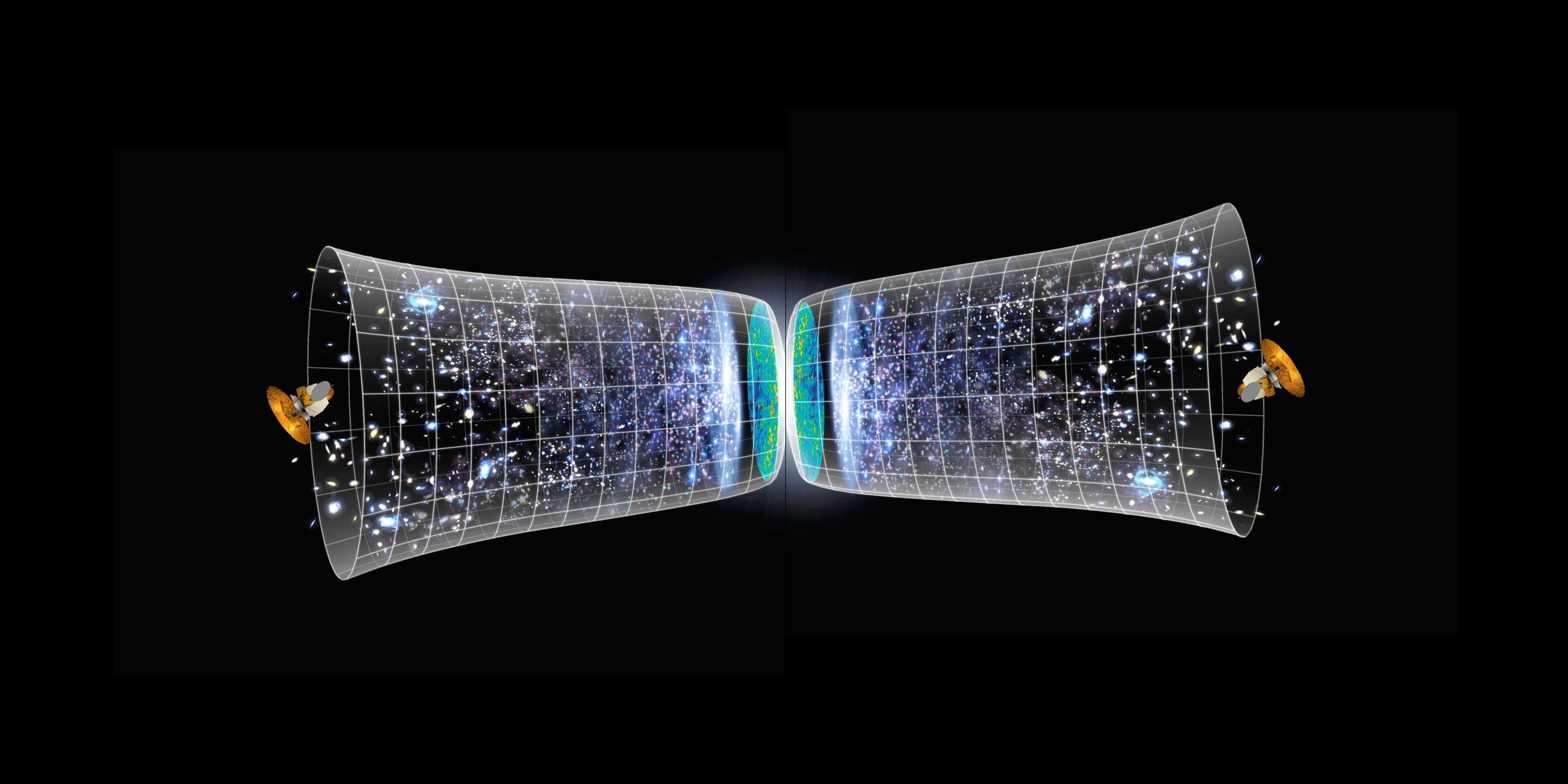}   \label{ESO1} \hspace{0.5cm}
\includegraphics[width=70mm,height=45mm]{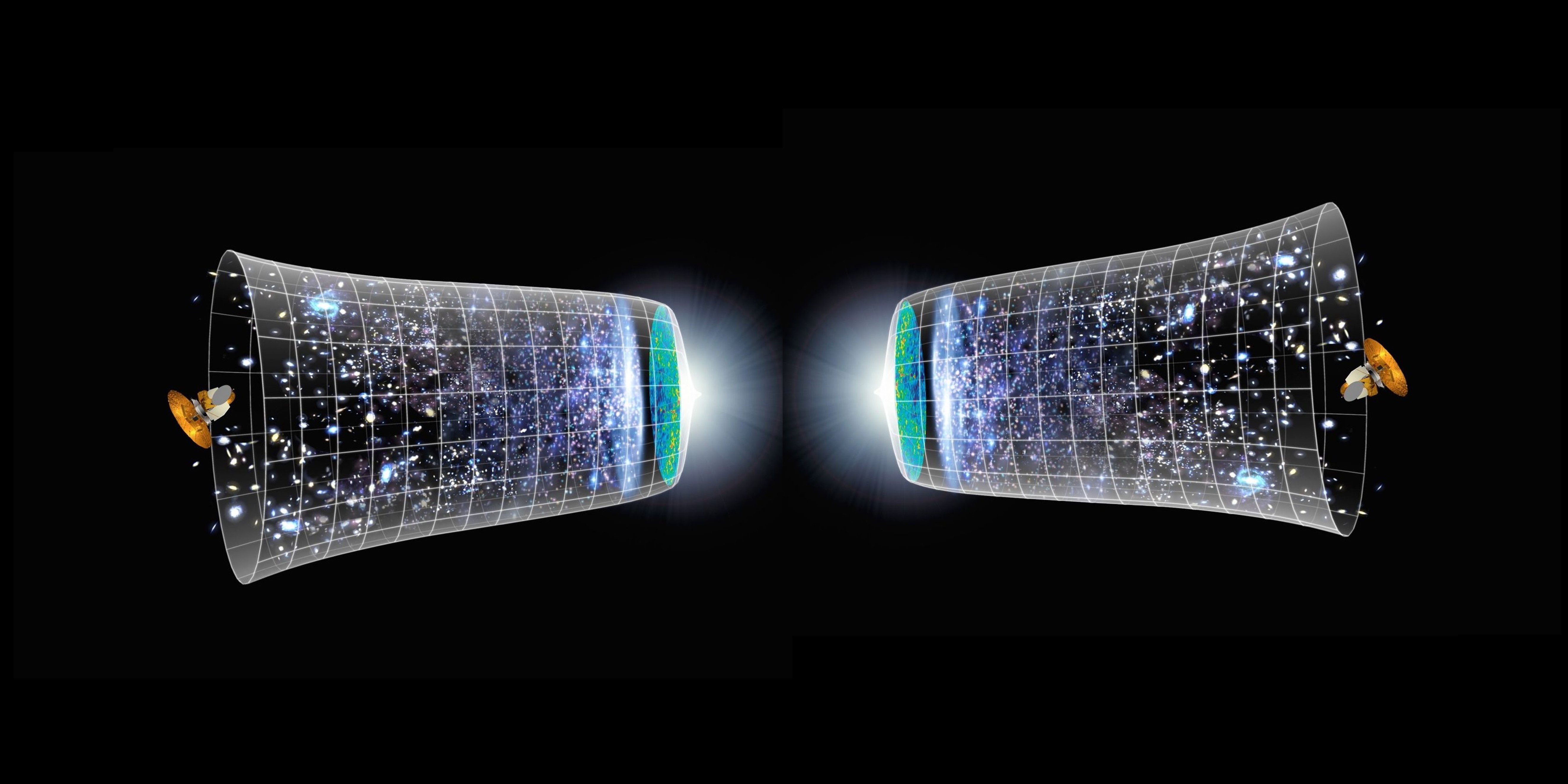}  \label{ESO2}
\caption{The figures show artistic representations of the cosmic contraction and expansion phases of the branch-cut universe evolution with two scenarios. In the representation sketched on the left figure, the branch-cut universe evolves from negative to positive values of the imaginary cosmological time $t_i$, circumventing continuously the branch-cut and no primordial singularity occurs, only branch points.  The right figure sketches the second scenario, where the branch cut and branch point disappear after the {\it realisation} of the imaginary time $t_i$ by means of a Wick rotation, which is replaced here by the real and continuous thermal time (temperature), $T$. In this second scenario, a mirrored parallel evolutionary universe, adjacent to ours, is nested in the structure of space and time, with its evolutionary process going backwards in the cosmological thermal time negative sector. The figures were based on artistic impressions~\cite{ESO}\label{ESO}.}
\end{figure*}

Before we conclude this contribution, we summarize below the present stage of the scenarios recently outlined~\citep{Zen2023b} for the branch-cut cosmology which are
sketched in an artistic representation 
(see Fig. (\ref{ESO})), with a branch-point and a branch-cut on the left figure, and no primordial singularity on the right one\footnote{Figures based on an artistic impression originally developed by ESO / M.~\citet{ESO}.}.

Two scenarios have been outlined in the branch-cut cosmology.
The first scenario is characterized by a branch-point and a branch-cut, with the suppression of a primordial singularity, in which the universe continuously evolves from the negative imaginary cosmological-time sector, --- 
where the contraction of the universe takes place ---,
to the positive one, ---  where the expansion of the universe occurs~\citep{Zen2020,Zen2021a,Zen2021b,Zen2022}.  This scenario, due to the absence of a singularity, may be characterized by a transition region with dimensions that surpass Planck's dimensions according to Bekenstein Criterion, separating the contraction and expansion phases~\citep{Zen2023a}. 

In the second scenario, the branch-cut and the branch-point disappear after realization of imaginary time through a Wick rotation, which is replaced here by the real and continuous thermal time (temperature)~\citep{Zen2020,Zen2021a,Zen2021b,Zen2022}. In this scenario, 
 the connection between the negative and positive imaginary cosmological-time sectors is {\it broken} as a result of the Wick rotation, and a mirrored parallel evolutionary universe adjacent to our own is nested in the fabric of space and time, with its evolutionary process oriented in the opposite direction to that corresponding to the positive imaginary-time sector~\citep{Zen2020,Zen2021a,Zen2021b,Zen2022}. 
 
 Singularity means there is no way for spacetime to start smoothly. Branch-cut cosmology, alternatively, proposes an absolutely non-temporal beginning in the imaginary sector, a configuration of pure space, through a Wick rotation that replaces the imaginary component of time with the temperature, the thermal time, that flows in the opposite direction of the arrow of time in the expansion phases of the first and second scenarios.
In the first scenario, in the contraction phase, before entering into the expansion phase, the temperature and entropy of the branch-cut universe must reach values consistent 
with the corresponding ones in the expansion phase. For this to happen, the temperature of the universe in the contraction phase must increase, but the entropy must decrease, as determined by thermodynamics, reversing this way the arrow of time.  
 In the contraction sector of the first scenario, as the transition region approaches, there occurs a progressive decrease in the entropy and an increase in the temperature of the universe, so there is a critical region, whose dimensions are determined by the Bekenstein Criterion~\citep{Zen2023a}, where the entropy reaches its minimum value and the temperature in contra-position its maximum value.
In the expansion sector, the opposite occurs where, having as a starting point the minimum values of entropy and maximum values of the temperature of the universe, progressively the entropy starts  to increase and the temperature to decrease. In this sense, in the contraction region, the arrow of time points in the opposite direction of the evolving orientation of the universe, while in the expansion region it points in the same direction. On the other hand, thermal time points in the evolutionary direction of the universe in the region of contraction and in the opposite direction in the region of expansion. 

Going back to the original conception, the arrow of time is associated with the direction of entropy increase and therefore oriented towards the expansion of the universe. However, in the contraction phase of the universe, entropy decreases and the arrow of time therefore has, in this conception, its direction inverted. One way to overcome such a discrepancy in order to maintain the ``past-to-future''  global orientation in the branch-cut cosmology is to define a time arrow oriented towards the decreasing of entropy in the contraction sector of the universe, and the conventional conception in the expansion phase. This proposition would be valid for both scenarios of the branch-cut cosmology, as in the first scenario there is a contraction sector followed by an expansion phase of the universe, while in the second scenario all sectors are associated with the expansion of both, ours and the mirror-universe. In this conception, the cosmological arrow of time is determined as the direction in which ``time'', from the macroscopic point of view, flows globally. This conception differs from that of George Ellis~\citep{Ellis, Ellis2014}, who identifies the arrow of time with the locally determined direction in which time flows at any time in the evolution of the universe. In the branch-cut cosmology, in turn, this local component is identified with the corresponding real component of the complexified-time that results from the complexification of the FWLR metric~\citep{Zen2021a}.

%%%%%%%%%%%%%%%%%%%%%%%%%

\section{Final Remarks}

The limitations presented by conventional cosmology arise not only from the fact that the ratio of the horizon size to the patch size shrinks when extrapolating back in time but essentially from the existence of singularities that break down this ratio and thus make it impossible to restore causality. In the branch-cut cosmology, this problem is overcome by the presence of cuts, defining this way the range of $\ln^{-1}[\beta(t)]$, making the scale factor shrinks as stressed before to a finite critical size. 
Furthermore, the present formulation assumes the description of the background evolution of the branch-cut cosmological scenarios in leading order by classical equations of motion.

The anisotropy associated with the smoothness problem of small temperature fluctuations has a peculiar solution that still lacks a more consistent explanation. This is because, in the final phase of the contraction process, choosing the epsilon parameter with a negative value (dark matter?), the scaling behavior of the ratio  $\frac{a^{2\epsilon_{DM}}(t)}{a^6(t)}$ is reversed, simultaneously assuming the remaining ratios $\frac{a^{\epsilon}(t)}{a(t)}$ are locked in. 

This finite size limitation of the range, well above the Planck limit,  implies in the branch-cut cosmology~\citep{Zen2023a} a leap, of a classical nature, between the primordial phases of contraction and expansion, a topic that needs further investigation in the future. A quantum approach, more in line with the nature of this evolutionary domain, following a recent proposal~\cite{Zen2021b}, is under investigation. Other aspects to be investigated refer to a more consistent evolution mapping of the cut-planes, the consequences of adopting a non-symmetric approach regarding the dimensionless thermodynamics connection $\epsilon(t)$, a different ordering of values of $\epsilon(t)$, and the role of dark matter in the evolution of the branch-cut universe. Cosmic Microwave Background (CMB) measurements by the Planck satellite~\cite{Planck} offer on the other hand an unprecedented opportunity to constrain and test the branch-cut model, particularly regarding the fluctuations associated to the energy density primordial spectrum, the seeds of all structures in the early universe. There are also questions regarding the multiverse content. For example, the conjuncture of a thermal contact of primordial multi-universes, immersed in a thermal bath, subjected to a contraction crunch, and the thermal evolution consequences to our universe after decoupling. And the ensuing questions, as the homogeneity of the primordial radiation and average density of matter as key elements for the formation of the structures observed today in the univers, among others. Evidently, most of these questions should not be limited to a classical description of the evolutionary universe, as their nature is intrinsically quantum. These topics are presently under investigation.
%

%%%%%%%%%%%%%%%%%%%%%%%%%

\bibliography{Zen}

\end{document}